\documentclass[aps,prl,superscriptaddress,twocolumn,showpacs,natbib]{revtex4}
\usepackage[english]{babel}
\usepackage{color}
\usepackage{graphicx}
\usepackage{cancel}
\usepackage{bm}
\usepackage{amsmath}
\usepackage{amssymb}

\def  \bsig    {\mbox{\boldmath$\sigma$}}
\def  \bta     {\mbox{\boldmath$\tau$}}

\begin{document}
\title{Enhanced photogalvanic effect in graphene due to Rashba spin-orbit coupling}

\author{M. Inglot}
\affiliation{Department of Physics, Rzesz\'ow University of Technology,
al.~Powsta\'nc\'ow Warszawy 6, 35-959 Rzesz\'ow, Poland}

\author{V. K. Dugaev}
\affiliation{Department of Physics, Rzesz\'ow University of Technology,
al.~Powsta\'nc\'ow Warszawy 6, 35-959 Rzesz\'ow, Poland}
\affiliation{Departamento de F\'isica and CeFEMA, Instituto Superior T\'ecnico,
Universidade de Lisboa, av.~Rovisco Pais, 1049-001 Lisbon, Portugal}

\author{E. Ya. Sherman}
\affiliation{Department of Physical Chemistry, Universidad del Pa\'is Vasco UPV-EHU, 48080 Bilbao, Spain}
\affiliation{IKERBASQUE Basque Foundation for Science, Bilbao, Spain}

\author{J. Barna\'s}
\affiliation{Faculty of Physics, Adam Mickiewicz University, ul. Umultowska 85,
61-614 Pozna\'n, Poland}
\affiliation{ The Nano-Bio-Medical Centre, Umultowska 85, 61-614 Pozna\'n, Poland}

\begin{abstract}
We analyze theoretically optical generation of a spin-polarized charge current (photogalvanic effect)
and spin polarization  in
graphene with Rashba spin-orbit coupling. An external magnetic field is applied in the graphene plane, which plays a crucial role in the
mechanism of current generation. We predict a highly efficient resonant-like photogalvanic effect
in a narrow frequency range which is determined by the magnetic field. A relatively less
efficient photogalvanic effect appears in a
broader frequency range, determined by the electron concentration and spin-orbit coupling strength.
\end{abstract}

\date{\today}

\maketitle

\textit{Introduction:}\;
Two-dimensional electron systems with spin-orbit (SO) coupling are currently of a broad interest due to the coupled charge and
 spin dynamics, as revealed in a variety of
spin related transport phenomena~\cite{dyakonov2008,sinova2014,aronov89,edelstein90}.
Owing to the SO coupling, the spin dynamics can be generated, among others,  by a low-frequency
electric field~\cite{rashba2003} as well as optically by interband electronic transitions~\cite{stevens2003,kato04,silov04}.
Moreover, an external static magnetic field can enable the current generation by light absorption,
leading to a photogalvanic effect~\cite{ganichev2003}.

Many of the spin related phenomena, including also the ones mentioned above, can be observed in two dimensional graphene monolayers
and other graphene-like materials like silicene for instance. The huge interest in
graphene is related mainly to its natural two-dimensionality, very unusual electronic
structure, and high electron mobility which ensures its excellent transport properties~\cite{GeimNatMater07,katsnelson}.
Even though the intrinsic SO interaction in free-standing graphene is negligibly small,
the Rashba spin-orbit coupling can be rather strong for graphene deposited
on certain heavy-element substrates~\cite{dedkov08,varykhalov08,zarea09}.
Since the electronic band structure of graphene is significantly different from that of a
simple two-dimensional electron gas (2DEG), and the spin-orbit coupling creates a gap
in the electronic spectrum, graphene can reveal qualitatively new effects which can not be observed in 2DEG in
conventional semiconductor heterostructures.
Additionally, the SO-related phenomena in graphene are also important from the point of view of potential applications
in all-graphene based
spintronics devices~\cite{trauzettel07,han2014,seneor2012}.

In this letter we predict an enhanced photogalvanic effect in graphene. To do this we consider the charge and spin currents generated
in graphene by optical pumping in the infrared photon energy region, and show that the optical pumping can be used to generate in graphene not only the
spin density~\cite{inglot14}, but also a spin-polarized
net current. An external magnetic field applied in the graphene plane plays an important role
in the mechanism of current generation.
We show that the efficiency of current generation per absorbed photon can be very high at certain conditions. Apart from this, we also show that one can
create spin density without creating electric current, but not {\it vice versa}.


\textit{Model:}\;  We consider low-energy electronic spectrum of graphene
in the vicinity of the Dirac points~\cite{katsnelson}. Additionally, we include
the Rashba SO coupling~\cite{kane05} and the Zeeman energy in a weak external in-plane magnetic field ${\bf B}$.
The corresponding Hamiltonian can be then written in the form
\begin{eqnarray}
\label{1}
\hat{H} =\hbar v_{0}(\pm \tau _xk_x+\tau _yk_y) +
\lambda (\pm \tau_x\sigma _y-\tau _y\sigma _x)
+\frac{\Delta}{2}({\bf b}\cdot \bsig),
\end{eqnarray}
where $v_{0}\simeq 10^{8}$cm/s is the electron velocity in graphene, $\tau _x$ and $\tau _y$ are the Pauli matrices defined in the sublattice space,
${\Delta}\equiv\,gB$ is the maximum Zeeman splitting, ${\bf b}\equiv {\bf B}/B,$
while the $+$ and $-$ signs refer to the $K$ and $K'$ Dirac points, respectively.
Furthermore,  $g=g_L\mu _B$  with $g_L=2$ being the Land\'e factor, and
$\lambda =\alpha/2$ with $\alpha$ standing for the Rashba coupling constant~\cite{rashba09}.

\begin{figure}
  \includegraphics[width=.9\columnwidth]{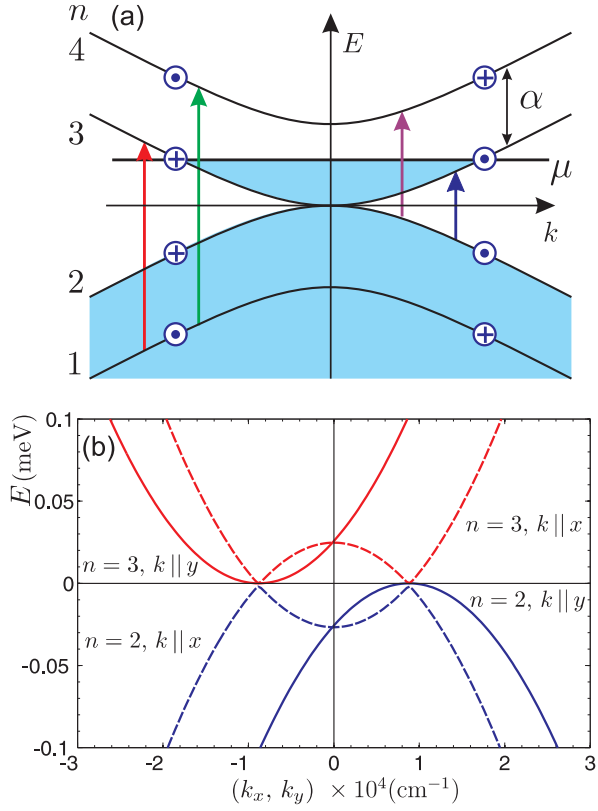}
  \caption{\label{fig1}(Color online). (a) Schematic picture of the band structure
  of graphene with Rashba spin-orbit interaction with all possible band
  transitions for a chosen chemical potential, indicated by the vertical arrows.
	Circles with crosses and dots
inside correspond to the opposite spin orientations in the subbands.
  (b) Dispersion of the low energy states for indicated wavevector orientations.
	The band index is also marked at the plots.
   The assumed magnetic field ${\bf B}\parallel x$ is equal to 5 T.}
\end{figure}

The electronic spectrum corresponding to Hamiltonian (\ref{1}) consists of four
energy bands in each valley, $E_{n{\bf k}}$, where $n=1$ to $n=4$ is the band index.
The corresponding spectrum for  $B=0$ is shown in Fig.\ref{fig1}a, and is
given by the formula $E_{n,{\bf k}}(B=0)=\mp \lambda \pm \sqrt{\lambda ^{2}+\hbar^{2}v_{0}^{2}k^{2}}$.
In turn,  the expectation value of the spin in the absence of magnetic field is oriented
perpendicularly to the wavevector
${\bf k}$ \cite{rashba09}, similarly as in a semiconductor-based 2DEG with Rashba SO coupling.
Contrary to the 2DEG, there are, however, no eigenstates of Hamiltonian (\ref{1}) with
a definite eigenvalue of
any spin component. This appears due to the specific form of the  SO coupling in graphene.
The corresponding spin components in the absence of magnetic field and for the
wavevector ${\mathbf k}\equiv k\left(\cos\theta,\sin\theta\right)$ are
\begin{subequations}
\begin{alignat}{2}
\langle \sigma_{x}\left({\mathbf k}\right)\rangle_{B=0} = \mp\frac{v_{k}}{v_{0}}\sin\theta, \label{sigx} \\
\langle \sigma_{y}\left({\mathbf k}\right)\rangle_{B=0} = \pm\frac{v_{k}}{v_{0}}\cos\theta, \label{sigy}
\end{alignat}
\end{subequations}
where $v_{k}=v_{0}\times{\hbar\,v_{0}k}/{\sqrt{\lambda ^{2}+\hbar^{2}v_{0}^{2}k^{2}}}$ is the
absolute value of the
electron velocity in the absence of magnetic field. Here the upper and lower signs correspond to
the bands (1,4) and (2,3), respectively.
Note, both spin components given by Eqs.~(\ref{sigx}) and (\ref{sigy}) vanish in the limit of
$k=0$~\cite{rashba09} corresponding to
the mixed rather than pure character of the band states in the spin subspace.

The electronic spectrum presented in Fig.\ref{fig1}a is significantly modified by an
external in-plane magnetic field.
Assume this field is oriented along the axis $x$. The exact electronic spectrum can be then
obtained by direct
diagonalization of the Hamiltonian
(\ref{1}), and is shown in Fig.\ref{1}b for wavevectors along the axis $x$ and $y$. Only the states corresponding to the bands
labeled in Fig.\ref{fig1}a  with the index 2 and 3 are shown there.  As one can note, for one propagation orientation
the electron bands are shifted vertically, i.e. to higher (lower) energy, while for the second propagation orientation the
bands are shifted horizontally, i.e. to left (right) from the point $k=0$. The latter separation in
the {\bf k}-space of the bands 2 and 3 is crucial for the enhanced photogalvanic effect.
For a weak Zeeman energy, $\Delta\ll\alpha$, and for electron momenta of our
interest, $k\gg\sqrt{\alpha\Delta}/\hbar v_{0}$,
the field-dependent correction to the electron energy, calculated by perturbation theory, has the form
\begin{equation}
E_{n,k}(B)-E_{n,k}(B=0)=\mp\frac{\Delta}{2}\frac{v_{k}}{v_{0}}\sin \theta ,
\label{correction}
\end{equation}
where again the upper (lower) sign corresponds to the bands (1,4) and (2,3), respectively.

Assume now that the system is subject to electromagnetic irradiation. Hamiltonian describing interaction of electrons in graphene with the external
periodic electromagnetic field,
$\mathbf{A}(t)=\mathbf{A}_0e^{-i\omega t}$, takes the form
\begin{eqnarray}
\label{2}
\hat{H}_A=\mp \frac{e}{c}v_{0}\, (\bta\cdot {\bf A}).
\end{eqnarray}
As in Eq. (\ref{1}), different signs correspond  here to electrons within the $K$ and $K'$ valleys.

The injection rate of a quantity ${\cal O}$, related to the intersubband optical transitions,
can be calculated by using the Fermi's golden rule,
\begin{eqnarray}
\label{3}
&&{\cal O}(\omega) =\sum _{n,n^{\prime}} {\cal O}^{n\rightarrow n^{\prime}}(\omega), \nonumber \\
&&{\cal O}^{n\rightarrow n^{\prime}}(\omega) =\frac{2\pi }{\hbar }
\int \frac{d^{2}{\bf k}}{(2\pi )^{2}}\;
\left|\langle\Psi_{n{\bf k}}|\hat{H}_A|\Psi_{n^{\prime}{\bf k}}\rangle
\right|^2 \widehat{O}^{n\rightarrow n^{\prime}}
\nonumber\\
&&\times\delta (E_{n{\bf k}}+\hbar\omega - E_{n^{\prime}{\bf k}})
f(E_{n{\bf k}})\,[1-f(E_{n^{\prime }{\bf k}})],
\end{eqnarray}
where $f(E_{n{\bf k}})$ is the Fermi-Dirac
distribution function. Since there are two valleys, $K$ and $K'$, one needs to
calculate contributions to ${\cal O}(\omega)$ from both of them.
The quantities of our interest here are:
\begin{equation}
\widehat{O}^{n\rightarrow n^{\prime}}\equiv \widehat{\openone},
\end{equation}
for the light absorption (where $\widehat{\openone}$ is the identity operator),
\begin{equation}
\widehat{O}^{n\rightarrow n^{\prime}}\equiv \langle
\Psi_{n^{\prime}{\bf k}}|\sigma _\nu|\Psi_{n^{\prime}{\bf k}} \rangle
-\langle \Psi_{n{\bf k}}|\sigma _\nu |\Psi_{n{\bf k}} \rangle,
\end{equation}
for the corresponding spin component injection, and
\begin{equation}
\widehat{O}^{n\rightarrow n^{\prime}}\equiv \langle
\Psi_{n^{\prime}{\bf k}}|\hat I_i|\Psi_{n^{\prime}{\bf k}} \rangle
-\langle \Psi_{n{\bf k}}|\hat I_i|\Psi_{n{\bf k}} \rangle
\end{equation}
for the current injection. Here, $\hat I_i\equiv e\hat v_i$, where $e$ is the electron charge,
while $\hat v_{x}\equiv\pm v_{0}\tau_{x}$ and $\hat v_{y}\equiv v_{0}\tau_{y}$.
The injected spin current, in turn, can be calculated as \cite{rioux14}
\begin{equation}
\widehat{O}^{n\rightarrow n^{\prime}}\equiv \langle
\Psi_{n^{\prime}{\bf k}}|\hat J_i^{\nu}|\Psi_{n^{\prime}{\bf k}} \rangle
-\langle \Psi_{n{\bf k}}|\hat J_i^{\nu}|\Psi_{n{\bf k}} \rangle,
\label{spincurrent}
\end{equation}
where $\hat J_{i}^{\nu}=[\sigma_\nu,\hat v_i]_{+}/{2}$.
Below we concentrate on the results for injection of charge current and spin density.

\begin{figure}%
\includegraphics[width=0.9\columnwidth]{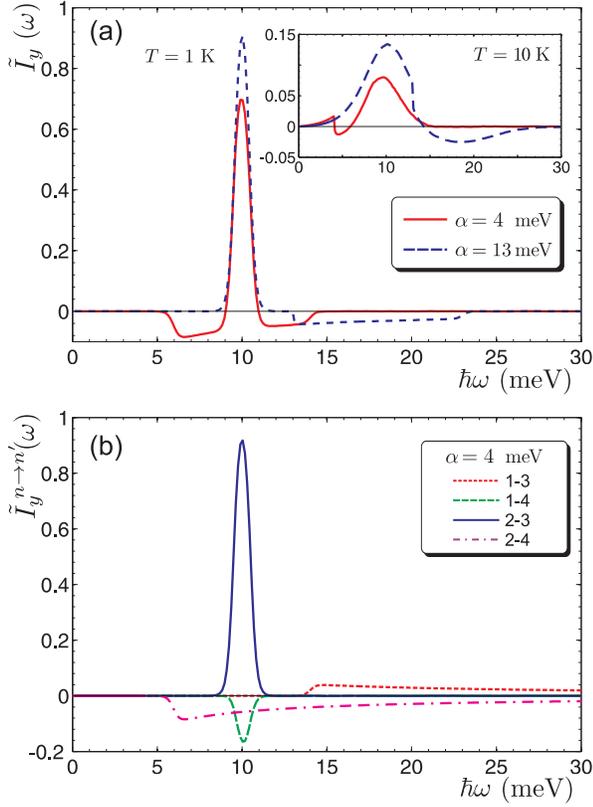}
\caption{\label{fig2}(Color online) (a) Normalized charge current $\tilde{I}_y$, in
the case of low temperature, $T=1$ K  ($T=10$ K in the inset).
Rashba spin-orbit coupling strength is $\alpha=4$ meV (solid red line)
and $\alpha=13$ meV (dashed blue line). The chemical potential is $\mu=5$ meV, $\bf B$ is
in the plane of graphene and along the $x$-axis, while $\bf A\, ||\, \bf B$.
(b) Contributions of indicated intersubband transitions to the total current presented in (a)
for $\alpha=4$ meV.}
\end{figure}

\begin{figure}%
\includegraphics[width=0.9\columnwidth]{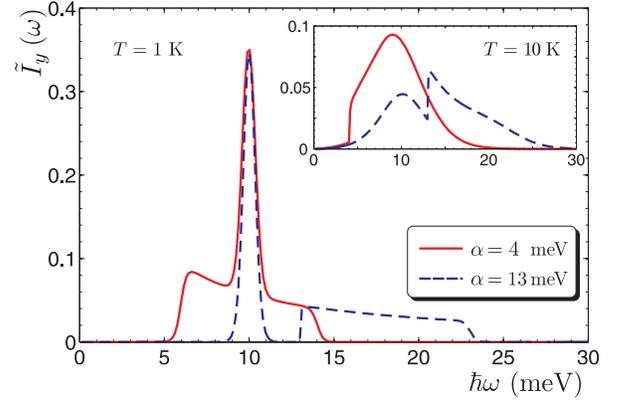}%
\caption{\label{fig3}(Color online) Normalized charge current $\tilde{I}_y$,
for a low temperature, $T=1$ K ($T=10$ K in the inset).
Solid red line is for $\alpha=4$ meV and
dashed blue line is for $\alpha=13$ meV. Chemical
potential $\mu=5$ meV, $\bf B$ is along the $x$-axis and $\bf A\, \bot\, \bf B$.}
\end{figure}

\textit{Results:}\;  Using equations for the injection rate one
can calculate the charge current and spin polarization
induced by the optical pumping.
Let us begin with the photogalvanic effect, i.e. charge current generation. Results for two
different polarizations of the electromagnetic field ${\bf A}(t)$ are
presented in Figs. \ref{fig2} and \ref{fig3}. Here the injection efficiency $\tilde{I}_{i}$
is defined as $\tilde{I}_{i}\equiv I_{i}/ev_{0}\mathcal{I}_0$,
where $\mathcal{I}_0=\pi e^2Q/\hbar c$ and $Q$ is the incident photon flux \cite{gusynin06,kuzmenko08,zulicke10}.

\begin{figure}
  \includegraphics[width=.9\columnwidth]{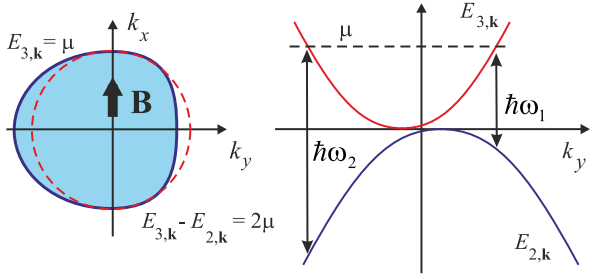}
  \caption{\label{fig4}(Color online). Left: Schematic picture of the Fermi line (solid) and resonance line
  (dashed) for the chosen subbands. Optical transitions are possible only at the part of the dashed line
  outside the filled area. Each transition generates a current of the order of $2ev_{k_{F}}$,
  making the generation highly efficient. Right: Side view on the intersubband transitions,
	which can occur in the   frequency interval $\omega_{1}\le\omega\le\omega_{2}$.}
\end{figure}

The transitions start at $\hbar\omega\approx 2\mu-\alpha$ if $\mu\ge\alpha$ and at $\hbar\omega\approx 2\mu$
otherwise. In both cases a strong narrow peak in the injection efficiency appears at a resonant energy
$\hbar\omega\approx 2\mu$. This peak is remarkably higher for the electromagnetic field polarized along the static magnetic field $\bf B$,
compare Figs. \ref{fig2} and \ref{fig3}. To understand origin of this peak lat us consider
transitions between the subbands marked with $n=2$ and $n=3$. First, we determine the shape of
the isoenergetical line
corresponding to a given Fermi energy, $\mu\gg \Delta$. For the band corresponding to $n=3$, one obtains from
Eq. (\ref{correction}) the first-order correction to the Fermi wavevector,
\begin{equation}
\hbar k_{F} =\frac{\sqrt{\mu ^{2}+2\lambda \mu }}{v_{0}}\mp \frac{\Delta}{2v_{0}}\sin \theta ,
\end{equation}
which sets the following boundaries for the Fermi surface:
\begin{subequations}
\begin{alignat}{2}
&-\frac{\sqrt{\mu ^{2}+2\lambda \mu }}{v_{0}}-\frac{\Delta}{2v_{0}} <
\hbar k_{F,y}<\frac{\sqrt{\mu ^{2}+2\lambda \mu }}{v_{0}}-\frac{\Delta}{2v_{0}} \label{FSshapea} \\
&-\frac{\sqrt{\mu ^{2}+2\lambda \mu }}{v_{0}}<\hbar k_{F,x}<\frac{\sqrt{\mu^{2}+2\lambda\mu}}{v_{0}}.
\label{FSshapeb}
\end{alignat}
\end{subequations}
Due to the ${\bf k}$-dependent Zeeman term, the Fermi surface becomes considerably deformed
and anisotropic, as shown
in the left panel of Fig.\ref{fig4}. The maximum deformation is independent of the chemical
potential and spin-orbit coupling. In turn, the resonance line
determined by $E_{3,\bf{k}}-E_{2,\bf{k}}=\hbar\omega$ is still a circle given by
the condition
\begin{equation}
\hbar k_{\omega }=\frac{\sqrt{\hbar^{2}\omega^{2}/4+\lambda\hbar\omega}}{v_{0}}.
\end{equation}
A part of the resonance line is inside the occupied region.
Therefore, we have an interval of the photon energies, $\left(\hbar\omega_{1},\hbar\omega_{2}\right)$,
as shown in the right panel of Fig.\ref{fig4}, where the transitions occur for positive values of
$k_y$, while the transitions with negative $k_y$ (which would compensate partly current) are
forbidden. As a result, a very efficient current injection
occurs in this photon energy window, as visible in Fig. \ref{fig2}.
In the considered regime of $\mu\gg\Delta$, this photon energy interval is determined by the conditions
\begin{subequations}
\begin{alignat}{2}
&\hbar\omega_{1}=2\mu-{\Delta}\frac{v_{k_{F}}}{v_{0}}, \\
&\hbar\omega_{2}=2\mu+{\Delta}\frac{v_{k_{F}}}{v_{0}},
\end{alignat}
\end{subequations}
which results in the peak width given by the formula,
\begin{equation}
\hbar\left(\omega_{2}-\omega_{1}\right)=2\frac{\sqrt{\mu ^{2}+2\lambda \mu }}{\mu +\lambda }\Delta.
\end{equation}
With the increase in temperature to $T>\Delta$, this effect becomes
smeared out by thermal broadening of the Fermi distribution, and the injection rate decreases
as shown in the insets to Figs. \ref{fig2} and \ref{fig3}.

Similar arguments can be also applied to the
transitions between $n=1$ and $n=4$ subbands. As a  result, one gets a relatively small
negative peak in the injection rate at $\hbar\omega\approx 2\mu$,
see Fig. \ref{fig2} (b). The weakness of this injection channel is due to a relatively small Fermi
velocity in the subband
4 at $\mu-\alpha\ll\alpha$, while its reversed sign is due to the opposite spin orientation
in these subbands, which results (similar to Eq. (\ref{FSshapea}) and Fig. \ref{fig4}) in a different shape of the Fermi surface, where the transitions
begin to occur at
$k_{y}<0$. In the limit $\alpha\ll \mu$, the positive and negative contributions compensate each other,
and the current injection efficiency tends to zero, as expected in the absence of spin-orbit coupling.

Having discussed the strong peaks in the current injection rate, let us consider now briefly
the broad structure. It is formed by momentum
dependence of the matrix elements and velocity, and has the efficiency of the order of $\Delta/\alpha$.
The current injection stops at $\hbar\omega\approx 2\mu+\alpha$,
where the contributions due to transitions between different bands compensate each other.
We also mention that for ${\bf B}\parallel x$, the charge current has only the $y$-component
for both polarizations of the incident light.

\begin{figure}
  \includegraphics[width=0.87\columnwidth]{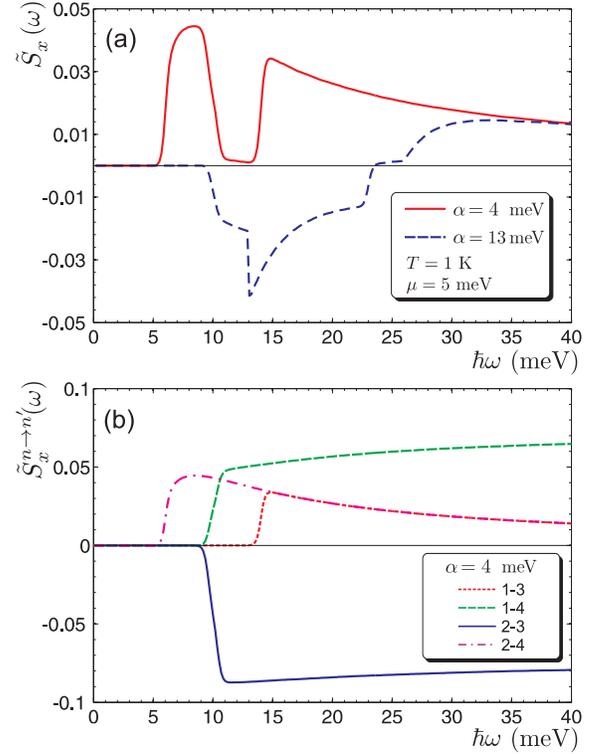}
	\caption{\label{fig5}(Color online).
  (a) Total injected normalized spin polarization $\tilde{S}_x=\sum_{n,n'}\tilde{S}_x^{n\to n'}$.
	Here the Rashba SO coupling $\alpha=4$ meV (solid red line) and $\alpha=13$ meV (dashed red line),
	$\mu=5$ meV and $B=5$~T. The orientations of ${\bf A}$ and ${\bf B}$ are parallel to the $x-$axis.
	(b) Transition-related spin injection $\tilde{S}_x^{n\to n'}$. }
\end{figure}

Now let us address briefly the problem of spin and spin current injection. For both incident
light polarizations one obtains a net spin polarization along the $x$ and $y$ axes.
The numerical results are presented in Fig. \ref{fig5} (a)
for the total spin polarization $S_x$, while Fig. \ref{fig5}(b) shows
injected spin polarization
associated with specific optical transitions. Physical mechanism of the  optically injected
spin polarization is rather clear,
since the spin-flip transitions are related to the above-mentioned fact that the
eigenstates of Hamiltonian (\ref{1}) are not the spin eigenstates, and the broken in magnetic field time-reversal symmetry allows one
to inject spin density.
Since the charge current  is along the $y$-axis, we obtain effectively a spin-polarized
current transferring in-plane spin components in the $y-$direction. As concerns the spin
current defined in Eq. (\ref{spincurrent}), it is
symmetric with respect to the time reversal and, therefore, magnetic field produces
there only changes proportional to $B^{2}$.

\textit{Summary:}\;
We have calculated optical injection  of charge current in graphene as the
photogalvanic effect due to spin-orbit coupling \cite{norris10}.
The current is injected only in a finite range of infrared light frequencies, determined
by the chemical potential $\mu$ and the spin-orbit coupling strength. The striking
feature of the injection is a narrow peak at the resonant frequency $\hbar\omega\approx 2\mu$, where
the current injection can be very efficient.
Comparing the $\omega$-dependence of the current and spin injection, we conclude that,
depending on the light frequency, one can inject either spin-polarized net electric current
or net spin polarization without the current injection.
This result can be applied to a controllable current generation in spin-orbit coupled graphene.

\textit{Acknowledgements.}
This work is supported by the National Science Center in Poland under Grant
No.~DEC-2012/06/M/ST3/00042. The work of MI is supported by the
project
No. POIG.01.04.00-18-101/12.
The work of EYS was supported by the University of Basque Country UPV/EHU under
program UFI 11/55, Spanish MEC (FIS2012-36673-C03-01), and ''Grupos
Consolidados UPV/EHU del Gobierno Vasco'' (IT-472-10).


%

\end{document}